\documentclass[11pt]{article}

\usepackage[round,sort,authoryear]{natbib}
\usepackage{authblk}

\usepackage{graphicx,amsmath,amssymb,fullpage,subfigure,url,xspace,color}
\usepackage{enumerat}

\newcommand{\blot}[1]{}

\usepackage{tikz}
\usetikzlibrary{arrows.meta,calc}

\tikzset{
  blk/.style={circle,fill=black,minimum size=6pt,inner sep=0pt},
  hol/.style={circle,draw,fill=white,minimum size=8pt,inner sep=0pt},
  holg/.style={hol,dash pattern=on 1pt off 1pt},
  edg/.style={line width=1pt},
  stalk/.style={dashed},
  stub/.style={dash dot},
  ar/.style={-{Stealth[length=3mm,width=2.2mm]},shorten >=3pt},
  fat/.style={ar,line width=1pt},
  blab/.style={blue},
  olab/.style={orange}
}

\newtheorem{theo}{Theorem}

\numberwithin{theo}{section}

\newcommand{\F}{{\mathcal F}}
\newcommand{\R}{{\mathbb R}}
\title{$\beta$-diversity and Graph Sheaf Laplacians}

\author{Peter Davidson\thanks{peter.davidson@strath.ac.uk}\quad}
\author{Michael Grinfeld\thanks{m.grinfeld@strath.ac.uk}}
\affil{Department of Mathematics and Statistics,\\
  University of Strathclyde, 26 Richmond Street,\\
  Glasgow, G1 1XH, UK}

\begin{document}

\maketitle

\begin{abstract}
  \noindent We suggest an approach to measuring biodiversity based on
  the spectrum of graph sheaf Laplacians associated with the sample
  data. In particular, we define a $\beta_{ss}$, a genuine
  generalisation of the multiplicative $\beta$-diversity. This scalar
  quantity is easily computable using methods of linear algebra. We
  show using simple examples that the notions we introduce (energies
  $E_k$ and $\beta_{ss}$) contain more information than just
  $\beta$-diversity.
\end{abstract}  

\medskip


\section{Introduction}

Whether we humans like it or not, we are the custodians of life on Earth. This is a heavy responsibility and we are slowly learning by trial and many a tragic error how to discharge it.

The first task is to be able to describe the sum total of life on Earth. Unfortunately, that task is a very hard one, the principles of which are unclear. A recent review of biodiversity \citet{Daly2018} is even entitled ``Ecological Diversity: Measuring the Unmeasurable''.

Below we simplify and equate diversity with richness, i.e., the {\em number} of species. More sophisticated approaches, for example using Hill's numbers to incorporate unevenness and similarity information \citep{Chao2016, Leinster2012} can be analysed in the framework suggested here, but we feel that explaining the principles we are using is best done in the simplest possible setting.

The point of our departure is that in ecological literature, one traditionally differentiates among $\alpha$-, $\gamma$-, and $\beta$-diversities. Of these $\alpha$ and $\gamma$ are relatively uncontroversial (once the term ``diversity'' itself has been defined): $\alpha$-diversity is defined in \cite{Daly2018} as referring to the ``diversity of a uniform habitat of a fixed size''; $\gamma$-diversity is ``the total diversity of the landscape''. However, $\beta$-diversity, which according to \cite{Daly2018}, ``refers to the average dissimilarity in composition among sub-communities'', is trickier and the literature on $\beta$-diversity is huge (e.g., see the discussions in \cite{Anderson2011,Baselga2010, Baselga2012, Baselga2015, Chao2019, Tuomisto2010}).

Simple examples show that the two widely accepted multiplicative \citep{Whittaker1972} and additive \citep{Lande1996} definitions of $\beta$-diversity are not very informative, as all distributions of species that agree on the average $\alpha$- and on $\gamma$-diversities lead to the same value of $\beta$-diversity.

Such cases differ in the way ``local becomes global''. So any $\beta$-diversity defined as a positive number obtained from directly manipulating $\gamma$- and $\alpha$-diversities will not have good discrimination properties as it will disregard adjacency relations of sampling sites.  We need to introduce an operator that takes the collection of $\alpha$-diversities and creates from them $\gamma$-diversity in a way that at the same time respects such adjacency relations, and thus is better able to differentiates between cases having the same $\beta$-diversity. In this contribution we demonstrate that such an operator can be constructed and that its spectrum (or trace) contains more information than the $\beta$-diversities accepted in the ecological literature.

The classical local-to-global tool is sheaf theory. The understanding that sheaf theory can be applied to data analysis dawned in about 2014--2015 in the work of Curry \citep{Curry2014} and Robinson \citep{Robinson2014, Robinson2017} and since then a considerable literature on this topic has appeared, mainly through the work of R. Ghrist and his students (e.g. \cite{Hansen2019, Hansen2020, Ghrist2020}). For a detailed recent review please see \citep{Ayzenberg2025} or the monograph \citep{Rosiak2022}. Other interesting contributions are to credence estimates \citep{Biesel2024} and to expert selection in the health services \citep{Aktas2025}. Sheaf theory is deep but computationally accessible, resting on rather simple linear algebra. Hence one can expect emerging applications of sheaf theory to be of use to researcher communities without expertise in that area of mathematics.

Accessible sources for what follows are the set of lectures by Essl \citep{Essl2023} and the preprint by Hansen \citep{Hansen2020a}, as well as \cite{Aktas2025} mentioned above,

\section{Defining a sheaf Laplacian}

In applying sheaf theory to construct suitable operators (Laplacians) a number of choices have to be made. Briefly, we start with a connected simple graph $G=(V,E)$, with vertex set $V$ and edge set $E$, and an assignment of data to all the vertices and edges. We equip every vertex $v$ and edge $e$ with vector spaces (stalks) $\F_v$, $\F_e$, respectively, containing relevant data, and for each vertex $v$ adjacent to an edge $e$ define restriction maps from $\F_v$ to $\F_e$, which we denote by
$\F_{v \rightarrow e}$. Finally, for any edge $e$ connecting vertices $v$ and $u$ we define the coboundary map $\delta$ by
\[
  (\delta)_e = \F_{v \rightarrow e} - \F_{u \rightarrow e}.
\]

This setup is summarised in Figure~\ref{setup}, where for both vertices and edges we have chosen the vector spaces $\R^m$ of some suitable dimension $m$. Note that if we choose $\R^{n_i}$ for the vertex $v_i$ and $\R^{n_{ij}}$ for the edge $e_{ij}$ incident to $v_i$, we must have that $n_i \geq n_{ij}$.

\begin{figure}
\begin{center}
\begin{tikzpicture}

%%%%%%%%%%%%%%%%%%%%%%%%%%%%%% base graph G
\node[blk] (a)  at ( 1.20,3.64) {};
\node[blk] (vi) at ( 2.86,2.64) {};
\node[blk] (b)  at ( 1.80,1.10) {};
\node[blk] (vj) at ( 5.74,3.00) {};
\node[blk] (c)  at ( 8.86,1.96) {};
\node[blk] (d)  at (10.16,3.28) {};

\draw[edg] (a) -- (vi) -- (b);
\draw[edg] (vi) -- (vj) -- (c) -- (d);

\node[below right=1pt] at (vi) {$v_i$};
\node[below=2pt] at (vj) {$v_j$};
\node at (6.14,1.04) {$G$};

%%%%%%%%%%%%%%%%%%%%%%%%%%%%%% stalks over the vertices
\node[hol] (ha)  at ( 1.26,5.84) {};
\node[hol] (hi)  at ( 2.90,4.94) {};
\node[hol] (hb1) at ( 1.80,2.70) {};   
\node[hol] (hj)  at ( 5.80,5.26) {};
\node[hol] (hc)  at ( 8.94,3.76) {};
\node[hol] (hd)  at (10.20,5.84) {};

\draw[stalk] (ha)  -- (a);
\draw[stalk] (hi)  -- (vi);
\draw[stalk] (hb1) -- (b);
\draw[stalk] (hj)  -- (vj);
\draw[stalk] (hc)  -- (c);
\draw[stalk] (hd)  -- (d);

%%%%%%%%%%%%%%%%%%%%%%%%%%%%%% stalks over the edges
\coordinate (mai) at ($(a)!0.48!(vi)$);
\coordinate (mib) at ($(vi)!0.62!(b)$);
\coordinate (mij) at ($(vi)!0.55!(vj)$);
\coordinate (mjc) at ($(vj)!0.53!(c)$);
\coordinate (mcd) at ($(c)!0.54!(d)$);

\node[hol] (sai) at ([yshift=9.4mm]mai) {};
\node[hol] (sib) at ([yshift=9.5mm]mib) {};
\node[hol] (eij) at ([yshift=9.6mm]mij) {};
\node[hol] (sjc) at ([yshift=9.5mm]mjc) {};
\node[hol] (scd) at ([yshift=8.3mm]mcd) {};

\draw[stub]  (sai) -- (mai);
\draw[stalk] (sib) -- (mib);
\draw[stalk] (eij) -- (mij);
\draw[stub]  (sjc) -- (mjc);
\draw[stub]  (scd) -- (mcd);

\node[below] at ([yshift=-2pt]mij) {$e_{ij}$};

%%%%%%%%%%%%%%%%%%%%%%%%%%%%%% restriction maps
\draw[fat] (hi) -- (eij);
\draw[fat] (hj) -- (eij);

\coordinate (mi) at ($(hi)!0.55!(eij)$);
\coordinate (mj) at ($(hj)!0.55!(eij)$);

\node (Fi) at (3.04,6.60) {$\mathcal{F}_{v_i\to e_{ij}}$};
\node (Fj) at (7.90,6.10) {$\mathcal{F}_{v_j\to e_{ij}}$};
\draw[ar] (Fi) to[out=0,in=110] (mi);
\draw[ar] (Fj) to[out=180,in=70] (mj);

%%%%%%%%%%%%%%%%%%%%%%%%%%%%%% annotations
\node at (2.50,5.24) {$\R^{n_i}$};
\node at (6.45,5.34) {$\R^{n_j}$};
\node (rij) at (3.84,0.70) {$\R^{n_{ij}}$};
\draw[ar] (rij) to[out=90,in=250] (eij);

\end{tikzpicture}
\end{center}
\caption{The basic set-up}\label{setup}
\end{figure}

The object we are after is the (graph) sheaf Laplacian, defined by 
\begin{equation}\label{lap}
  L_{G,D} = \delta^T \delta.
\end{equation}

Other choices of the underlying structure are possible.  Hypergraphs are considered in \cite{Duta2023}; alternatively, one could choose metric spaces of probability distributions instead of vector spaces for stalks, as is done in \cite{Biesel2024} or lattices of sets \citep{Ghrist2020}, but choosing vector spaces reduces computations to elementary linear algebra.

As we show below, the sheaf Laplacian we construct collects information both about the data and the structure of the graph of
sampling sites and so has greater expressivity than $\beta$-diversity.

The spectrum $\sigma_{L_{G,D}}$ of $L_{G,D}$ is real, and we call the sum of the $k$-th powers of eigenvalues the $k$-{\bf energy}, $E_k(L_{G,D})$, of the graph sheaf Laplacian:
\begin{equation}\label{ener}
  E_k(L_{G,D}) = \sum_{\lambda_i \in \sigma_{L_{G,D}}} \lambda_i^k .  
\end{equation}
Of course $E_1(L_{G,D})$ can be computed simply by evaluating the trace of $L_{G,D}$.

\subsection{Computing the Laplacian}\label{Lap}

Let us show in detail the computation of the energy for a path on three vertices. See \cite{Essl2023} and \cite{Hansen2020a} for more details of the theory. By a path we mean a graph $G$ on $n$ vertices in which $V=\{1,2,\ldots, n\}$ and $E= \{ 12, 23, \ldots, (n-1)n\}$, i.e., vertex $k$ ($1<k<n$) is connected to vertices $k-1$ and $k+1$.

Suppose we assign to vertex $v_1$ the data $\{a,b\}$, to vertex $v_2$ the data $\{a\}$ and to vertex $v_3$ the data $\{a,b,c\}$. Below for brevity in the case of a path we will summarise such an assignment to vertices as $\{ab,a,abc\}$. Let us also assign to both edges $e_{12}$ and $e_{23}$ the data $\{a\}$.

We associate with $v_1$ the vector space $\R^2$, with $v_2$ the space $\R$ and with $v_3$ the space $\R^3$. With the edge $e_{12}$ suppose we associate the space $\R$. Next we look for two linear maps into $\R$, one from $\R^2$ taking the vector $(a,b)^T \in \R^2$ from $v_1$ into the vector $a \in \R$, and one from $\R$ taking $a$ from $v_2$ into $a$. This means that associated with the edge $e_{12}$ we will have a row of the coboundary map $\delta$. Similarly, with the edge $e_{23}$ we associate the space $\R$ and look for two additional linear maps into $\R$, one from $\R$ into $\R$, taking the data of $v_2$, $a \in \R$, into $a \in \R$ and one from $\R^3$, taking the vector $(a,b,c)^T \in \R^3$ into $a \in \R$.

As there are $6$ (repeated) data values to be taken into account, the argument above means that the coboundary map $\delta$ will be a $2 \times 6$ matrix, and the corresponding Laplacian will be a $6 \times 6$ (symmetric) matrix.

Let us see how to construct the first row of $\delta$, which encodes the interaction of $v_1$ and $v_2$.  The first one,
$\F_{v_1 \rightarrow e_{12}}$, maps $\R^2$ into $\R$, taking the vector $(a,b)^T$ into $a$; clearly it is represented by the $1 \times 2$ matrix $[1 \quad 0]$; the second, $\F_{v_2 \rightarrow e_{12}}$, takes $\R$ into $\R$ mapping $a$ into $a$. It is represented by the $1 \times 1$ matrix $[1]$. Hence the first row of $\delta$ is the $1\times 6$ matrix
\[
  \left[\begin{array}{rrrrrr} 1 & 0 & -1 & 0 & 0 & 0 
  \end{array}\right],                                                 
\]
where the first two entries correspond to $\F_{v_1 \rightarrow e_{12}}$, the next one to $-\F_{v_2 \rightarrow e_{12}}$, and the remaining three zeros to $v_3$ (with which we associated $\R^3$).

Similarly, we encode the interaction of $v_2$ and $v_3$. As mentioned above, we associate with the edge $e_{23}$ the vector
space $\R$. $\F_{v_2 \rightarrow e_{23}}$ maps $\R$ into $\R$, taking $a$ into $a$, so it is again represented by the
$1 \times 1$ matrix $[1]$. $\F_{v_3 \rightarrow e_{23}}$ maps $\R^3$ into $\R$, taking the vector $(a,b,c)^T$ into $a$. Hence it is represented by
\[
  \left[\begin{array}{rrr} 1 & 0 & 0
        \end{array}\right].
\]
Putting all this information together, we have for the second row of $\delta$ the $1 \times 6$ matrix
 \[
    \left[\begin{array}{rrrrrr} 0 & 0 & 1 & -1 & 0 & 0
    \end{array}\right],                                                 
  \]
where the first two zeros correspond to $\R^2$ associated with $v_1$, the next entry is $\F_{v_2 \rightarrow e_{23}}$ and the last three come from $-\F_{v_3 \rightarrow e_{23}}$. Hence for $\{ab,a,abc\}$ the coboundary map is given by
\[
  \delta = \left[\begin{array}{rrrrrr}
               1 & 0 & -1 & 0 & 0 & 0 \\
               0 & 0 &  1 & -1 & 0 & 0
              \end{array}\right].                    
\]

  Once the coboundary map $\delta$ is constructed, the graph sheaf Laplacian is computed from (\ref{lap}), and the sum of its
  eigenvalues  gives the energy (\ref{ener}).

%%%%%%%%%%%%%%%%% Here Now %%%%%%%%%%%%%%%%%%%%%%%%%%%%%%%%%%%

\section{Constructions}

Above we were given a graph and an assignment of data to its vertices and edges, and constructed its Laplacian from that information. The question is now: How to use the same construction to produce the equivalent of $\beta$-diversity from sampling data? 

So suppose we have a sampling region with sites $1, \,\ldots, N$. At each site we are given the list of species found at the site. The adjacency matrix of sites will certainly define a graph $G_s$ (which we will call the {\em sampling graph}). We see a number of possibilities how to proceed further.

(a) We can use the sampling graph $G_s$ as our underlying graph, assign the sampling data to its vertices and assign to an edge the intersection of the data on the vertices joined by the edge. This is a natural construction, but unfortunately the resulting object is often not expressive enough.  It is easy to see in the following example: Let $G_s$ be a path on three vertices and two edges with data assignment $\{a,b,c\}$. Then the resulting Laplacian is easily seen to be the zero matrix. The construction of a graph Laplacian using directly the sampling graph $G_s$ and assigning intersection data to edges is also problematic because it is not clear how to deal with empty sites consistently.

We will discuss some non-trivial examples of this construction below.  

(b) A different idea is to construct a new graph, which we call $H(G_s)$, that coincides with the line graph of $G_s$ if $G_s$ has no vertices of degree $1$, assign data to its edges in a canonical way to be explained below and then work with its sheaf Laplacian. This approach turns out to be fruitful and allows for a genuine generalisation of the concept of $\beta$-diversity.

Thus, let the sampling graph $G_s$ have vertices $v_1, \ldots, v_N$, to which sampling data is assigned as above.  We proceed as follows:
\begin{enumerate}
  \item Associate with every vertex $v_j$ of $G_s$ of degree 1 a ``phantom vertex'' $w_j$ with empty set data and connect the two vertices by an edge $e_{jj}$. Denote this new graph by $G_s'$.
  \item Associate with each edge of $G_s'$ the union of data associated with the vertices of $G_s$ and phantom vertices.
  \item Define $H(G_s)$ to be the line graph, $L(G_s')$, of $G_s'$. Each vertex of $H(G_s)$ is assigned the data described in step 2. The edges are assigned the data originally associated to the vertices of $G_s$.
\end{enumerate}

Let us give an example of this construction. Suppose we start with the graph $G_s$ considered above, a path on three vertices $v_1, v_2, v_3$ with two edges $e_{12}$ and $e_{23}$ and data assignment $\{ab,a,abc\}$ for the vertices. First we enlarge the
vertex set to $w_1, v_1, v_2, v_3, w_3$. We assign to $w_1$ and $w_3$ empty set data. Then we assign to $e_{11}$ the data $\{a,b\} \cup \emptyset = \{a,b\}$, to $e_{12}$ the data $\{a,b\} \cup \{a\} = \{a,b\}$, to $e_{23}$ the data $ \{a \} \cup \{a,b,c\}= \{ a,b,c \}$ and to $e_{33}$ the data $\{a,b,c\} \cup \emptyset = \{a,b,c\}$. Thus the graph $H_G$ has 4 vertices, say, $u_1, \ldots, u_4$, to which we assign in order the data $\{ab, ab, abc, abc\}$, while to the edges, say $d_{12}$, $d_{23}$, $d_{34}$, we assign the data associated in the sampling graph $G_s$ with the vertices $v_1$, $v_2$, $v_3$, i.e., $\{ab,a,abc\}$ in that order. See Figure~\ref{hg}. Once the graph and data assignment are dealt with, we proceed as explained in Subsection~\ref{Lap}.

\begin{figure}
\begin{center}
\begin{tikzpicture}

%%%%%%%%%%%%%%%%%%%%%%%%%%%%%% top row : G_s
\node[holg] (t0) at (0,4) {};
\node[blk]  (t1) at (2,4) {};
\node[blk]  (t2) at (4,4) {};
\node[blk]  (t3) at (6,4) {};
\node[holg] (t4) at (8,4) {};

\draw[stalk] (t0) -- (t1);
\draw[edg]   (t1) -- (t2) -- (t3);
\draw[stalk] (t3) -- (t4);

\node[above=3pt] at (t0) {$\emptyset$};
\node[above=3pt] at (t4) {$\emptyset$};
\node[above=3pt] at (t1) {$S_1$};
\node[above=3pt] at (t2) {$S_2$};
\node[above=3pt] at (t3) {$S_3$};
\node at (4,5.6) {$G_s$};

%%%%%%%%%%%%%%%%%%%%%%%%%%%%%% bottom row : H(G_s)
\node[blk] (b0) at (1,2) {};
\node[blk] (b1) at (3,2) {};
\node[blk] (b2) at (5,2) {};
\node[blk] (b3) at (7,2) {};

\draw[edg] (b0) -- (b1) -- (b2) -- (b3);

\node[below=6pt] at (b0) {$S_1$};
\node[below=6pt] at (b1) {$S_1\cup S_2$};
\node[below=6pt] at (b2) {$S_2\cup S_3$};
\node[below=6pt] at (b3) {$S_3$};
\node at (4,-0.5) {$H(G_s)$};

%%%%%%%%%%%%%%%%%%%%%%%%%%%%%% vertical connectors (bottom nodes to top edges)
\draw[stub] (b0) -- (b0 |- t0);
\draw[stub] (b1) -- (b1 |- t0);
\draw[stub] (b2) -- (b2 |- t0);
\draw[stub] (b3) -- (b3 |- t0);

%%%%%%%%%%%%%%%%%%%%%%%%%%%%%% downward stubs (from the bottom edges)
\foreach \x/\lab in {2/S_1, 4/S_2, 6/S_3} {
  \draw[stub] (\x,2) -- (\x,0.8);
  \node[below] at (\x,0.8) {$\lab$};
}

\end{tikzpicture}
\end{center}
\caption{Example of the $H(G_s)$ construction. Here $S_1 = \{a,b\}$, $S_2 = \{a\}$ and $S_3 = \{a,b,c\}$.}\label{hg}
\end{figure}

Below we use the notation $L_{G_s,D}$ for the graph sheaf Laplacian of the sampling graph $G_s$ with data assignment $D$ as in (a) above and $L_{H(G_s),H(D)}$ for the graph sheaf Laplacian of the line graph $H(G_s)$ of $G_s$ with data assignment as in (b) above. Another construction we will introduce below is taking a graph $G$ with $N$ vertices, with an assignment of data $D$ on its vertices, and connecting all the vertices, thus creating a complete graph $K_N$. We can then assign to each of its edges $e_{ij}$ the intersection of the sets of data on $v_i$ and $v_j$. This data structure is denoted by $K(D)$ and its graph sheaf Laplacian by $L_{K(G), K(D)}$. Note that $H(K(G))$ as well as $L_{H(K(G)), H(K(D))}$ are well-defined and will be important below.
  
\section{Back to $\beta$-diversity: examples}

In this section we test the expressivity of our constructions using very simple examples, in most of which the habitat is composed of three sites. From now on we interpret the assignment of data such as $\{a,b\}$ to a vertex $v$ to mean that in site $v$ sampling found species $a$ and $b$. Then we can compute $\alpha$-, $\gamma$-, and $\beta$-diversities. For example, in the case of $\{ ab, ac, bc \}$ and of $\{ ab, a, abc \}$ (cases (\ref{i3}) and (\ref{i5}) below), $\gamma$-diversity is $3$ and the average $\alpha$-diversity is $2$, hence the multiplicative $\beta$-diversity is $1.5$ and the additive one is $1$ in both cases. But the situation clearly is different, for example, in terms of nestedness \citep{Baselga2010}, which is what has
motivated our work.

\subsection{Using the sampling graph $G_s$}

Now we present information from a number of simple path configurations with $\gamma$-diversity equal to $3$: $\sigma$ is the spectrum of $L_{G,D}$ with multiplicities of eigenvalues indicated by superscripts; we compute the $2$-energy $E_2$.
  
\begin{enumerate}[(i)]
\item $\{a,b,c\}$: $\sigma = \{ 0^3\}$, $E_2(G_s,D)=0$, $\beta=3$; \label{i1}   
\item $\{bc,a, abc\}$:  $\sigma =\{0^5,2\}$, $E_2(G_s,D)=4$, $\beta=2$;\label{i2}   
\item $\{ab,ac,bc\}$:  $\sigma=\{0^4,2^2\}$, $E_2(G_s,D)=8$, $\beta= 3/2$;
  \label{i3}   
\item $\{a,bc,abc\}$:  $\sigma=\{0^4, 2^2\}$, $E_2(G_s,D)=8$, $\beta=3/2$; \label{i4}   
\item $\{ab,a,abc\}$: $\sigma = \{0^4, 1,3\}$, $E_2(G_s,D)=10$,
  $\beta=3/2$; \label{i5}
\item $\{a,abc,bc\}$: $\sigma=\{0^3, 2^3\}$, $E_2(G_s,D)=12$, $\beta=3/2$; \label{i6}   
\item $\{a,ab,abc\}$: $\sigma=\{0^3,1,2,3\}$, $E_s(G_s,D)=14$,
  $\beta=3/2$; \label{i7}
\item $\{a,abc,ab\}$ : $\sigma=\{0^3,1,2,3\}$, $E_2(G_s,D)=14$,
  $\beta=3/2$; \label{i8}
\item $\{ab,ac,abc\}$ : $\sigma=\{0^4,1,2,3\}$, $E_2(G_s,D)=14$,
  $\beta=9/7$; \label{i9}
\item $\{ abc,abc, abc\}$: $\sigma=\{0^3, 1^3, 3^3\}$, $E_2(G_s,D)=30$,
  $\beta=1$.   \label{i10}
\end{enumerate}

We see that in the 6 cases considered that have multiplicative $\beta$-diversity of $3/2$, we have created $4$ different subclasses, with $E_2(G,D)=8$ ((\ref{i3})-(\ref{i4})), $E_2(G_s,D)=10$ (\ref{i5}), $E_1(G,D)=12$ (\ref{i6}) and $E_2(G_s,D)=14$ ((\ref{i7})-(\ref{i8})). In particular, we have separated (\ref{i3}) and (\ref{i5}).  (Note that using $E_1$ instead of $E_2$ will not discriminate between (\ref{i5}) and (\ref{i3}) and (\ref{i4})).

%%%%%%%%%%%%%%%%%%%%%%%% Here Now %%%%%%%%%%%%%%%%%%%%%%%%%%%%%%%%%%

In three cases the discrimination by $E_2$ is not sufficient: (\ref{i3})-(\ref{i4}), (\ref{i7})-(\ref{i8}), and (\ref{i8})-(\ref{i9}); in the latter case configurations with same energy have different values of $\beta$-diversity. We will comment on
this case below, but note that in the spectrum of (\ref{i8}) and (\ref{i9}) the zero eigenvalue has different multiplicity.

The case of (\ref{i8})-(\ref{i9}) can be dealt by considering $K(G_s)$ and computing the $2$-energy of $L_{K(G_s),K(D)}$. For example, cases (\ref{i5}), (\ref{i7}) and (\ref{i8}) correspond to the same configuration if adjacency is not taken into account. The computation of energy can be repeated. For the case of (\ref{i8}) we have $\sigma=\{0^3, 2, 3^2\}$, $E_2=22$ while for (\ref{i9}) we obtain $\sigma=\{0^3, 3^2,2^2\}$, $E_2=26$, thus separating the two cases.

Clearly (\ref{i7}) and (\ref{i8}) correspond to the same sheaf over $K(G_s)$, so cannot be separated. Interestingly, neither can (\ref{i3}) and (\ref{i4}), both of which over $K(G_s)$ have $2$-energy of $12$.

We note that the set-up suggested in this section leads to an interesting general problem in combinatorics: for all possible configurations of creating $n$ subsets (with repetition) from a set of $m$ elements, characterise the energy landscape and its discrimination ability.

To summarise: using notions of energy of the corresponding graph sheaf Laplacians allows us to discriminate between (some) cases that have the same average $\alpha$-, $\gamma$- and hence $\beta$- diversities. 

\section{Spatially sensitive $\beta$-diversity}

What is not clear from the above discussion is the direct connection
of the notions we have introduced and the classical multiplicative
$\beta$-diversity. In the present section we clarify this connection
and exhibit a quantity that genuinely generalises the classical
multiplicative $\beta$-diversity.

Let $G$ be any connected simple graph and let $D$ be some assignment of data to the vertices and edges of $G$. We let $D_{max}$ be the assignment of ``maximal possible data'' to the graph $G$, i.e., to each vertex and edge of $G$ we assign the union of all sets of data in $D$, and denote $E_1(L_{G,D_{max}})$ by $E_{1}^{max}(G,D)$.

We define the {\em spatially sensitive $\beta$-diversity}, $\beta_{ss}$, of a sampling graph $G_s$ with data assignment $D$ to be
\begin{equation} \label{bss}
  \beta_{ss}(G,D)=  \frac{E_1^{max}(H(G),H(D))}{E_1(H(G),H(D))}.
\end{equation}

This definition is a genuine generalisation of the classical multiplicative definition of $\beta$-diversity \citep{Whittaker1972} in the following sense.

\begin{theo}\label{paths_reg}
  If the sampling graph $G_s$ is a path or a $k$-regular graph $(k \geq 1)$ with vertex data assignment $D$, then
  \[
    \beta_{ss}(G_s,D) = \beta,
  \]
  the $\beta$-diversity of the sampling region covered by the sites in $G_s$.
\end{theo}  

We relegate the proof to the Appendix. Viewed one way, our definition of $\beta_{ss}$ in the case of paths and regular graphs simply provides a rather complicated way of computing the classical (multiplicative) $\beta$ diversity \citep{Whittaker1972}. However, this theorem shows in particular that if one forgets the adjacency relationships among sites in a sampling graph, and instead of $G_s$ considers $K(G_s)$, $\beta_{ss}$ reduces to $\beta$. Theorem~\ref{paths_reg} justifies calling our genuine generalisation of $\beta$-diversity ``spatially sensitive'', though perhaps ``regularity sensitive'' would also be appropriate.

A more general result, from which Theorem~\ref{paths_reg} follows and whose proof is also found in the Appendix, is given by

\begin{theo}\label{betass}
  For a sampling graph $G_s$ with data assignment $D$, we have
  \[
    \beta_{ss}(G_s, D) = \frac{\gamma}{r_c},
  \]
  where $\gamma$ is the $\gamma$-diversity of the sampling region covered by the sites in \(G_s\).
\end{theo}

Here $r_c$ is the {\em mean weighted richness} of the sampling graph,
defined in general as follows:
\[
  r_c = \frac{\sum_{v \in V} c(v)\alpha_v}{\sum_{v \in V} c(v)},
\]
where $c(v)$ depends only on the degree of vertex $v$ and $\alpha_v$
is the $\alpha$-diversity of this site in $G_s$. Below we take
$c(v) = \binom{d'(v)}{2}$, where $d'(v)$ is the degree of a vertex $v$
in $G_s'$. 

\medskip

We will now construct examples of a sampling graph with different
assignments of data to illustrate the role of $r_c$ in the computation
of $\beta_{ss}$. Let $G_s$ be a star graph with vertex set
$\{v, u_1, u_2, u_3, u_4\}$. The central vertex is labelled $v$ and
the leaf vertices are $u_1, \ldots, u_4$. Let $D_v, D_1, \ldots, D_4$
be the data assigned to these vertices.
\begin{enumerate}
\item Let $D_v =\{a, b, c, d\}$ and let $D_i = \{a\}$ for all $i$. In
  this case $r_c = \frac{28}{10}$ and $\beta_{ss} =
  \frac{20}{14}$. The multiplicative $\beta$-diversity is
  $\frac{5}{2}$ and so $\beta_{ss} < \beta$.
\item Let $D_v = \{a\}$ and let $D_i = \{a, b, c, d\}$ for all
    $i$. In this case $r_c = \frac{22}{10}$ and
    $\beta_{ss} = \frac{20}{11}$. The multiplicative $\beta$-diversity
    is $\frac{20}{17}$ and so $\beta_{ss} > \beta$.
\end{enumerate}

In case 1, there is a highly connected, rich central vertex that
dominates the sum $\sum c(v)\alpha_v$. This results in a larger value
of $r_c$, and hence a smaller value of $\beta_{ss}$. In case 2,
however, the smaller value of $r_c$ is a consequence of the sampling
graph having rich vertices of low connectivity.

\section{Discussion}

We have introduced a sheaf-theoretic approach to questions of
biodiversity, collecting statistics of species distribution over the
vertices of the sampling graph using energies $E_k$ and a
generalisation of the multiplicative $\beta$-diversity, $\beta_{ss}$.
Our framework seems to us to be ecologically informative and
mathematically rich, and we hope that it will be developed further by
both ecologists and mathematicians. Open questions abound and much
remains to be done: automating the computations for realistically
large graphs, leveraging to the present context the many tools used
with great success in network/graph theory, and understanding how to
combine the information obtained from the graph sheaf Laplacian
computations using different constructions.

One could also ask if once the graph sheaf is defined, there are other
ways to collect statistics of the data assignment apart from energies
and the resulting $\beta_{ss}$ which only rely on the spectrum of the
graph sheaf Laplacian.

Incidentally, one might speculate that in ecosystems energy decays in
time, and it would be interesting to see whether ecosystems minimise
the rate of decay of energy. Such a principle would have predictive
ability.

Even while concentrating only on the spectrum of the Laplacian, we did
not explore all the possible constructions. One suggestion is the
following. Adjoin to our graph $G_s$ on $N$ vertices an $(N+1)$-st
vertex to which we assign data containing all the species.  In other
words, we hard-wire $\gamma$-diversity into the new graph which we
will call $G_\gamma$, and connect this vertex with all the existing
vertices. Now compute $E_2$ of the sheaf Laplacian of the new
graph. This construction separates $\{ab, ac, abc\}$, the energy of
the graph sheaf Laplacian of $G_\gamma$ being $64$, and
$\{ad, ac, abc\}$, in which case the corresponding energy is
$66$. Unfortunately, this construction or still does not separate
(\ref{i3}) and (\ref{i4}), giving $E_2=46$ in both
cases. Understanding the difficulty here and separating these two
situations is left for further work.

An interesting construction we do not know how to implement is as
follows: given two vertices $v_i$ and $v_j$ of $G_s$, associate with
$e_{ij}$ the symmetric difference between the sets of species
contained in $v_i$ and $v_j$.

In the present paper, we always start with a graph and equip it with a
choice of vector spaces. It is interesting to consider when other
choices could lead to easily computable statistics, such as of metric
spaces or lattices instead of vector spaces, or of hypergraphs instead
of graphs, etc.

Ecological aspects of future work in the direction we suggest could
include how changes in biodiversity translate into, for example,
changes in energy, and how energy encodes different decompositions of
biodiversity, in particular of $\beta$-diversity
\citep{Baselga2010,Baselga2015}.

Above we equated biodiversity with richness. More sophisticated
approaches, such as those based on Hill's numbers
\citep{Chao2016,Chao2019, Leinster2012} are also left for future work.

To summarise: we have just started to understand what ecologically
important information is contained in the graph sheaf Laplacian and,
in particular, how it relates to and complements the information
provided by the classical $\beta$-diversity.

Finally, we expect that the approach of this paper has applicability
beyond the area of $\beta$-diversity. For example, the influential
work of Borgatti \citep{Borgatti2005} on network flows can be
interpreted in terms of sheaf dynamics.

\section{Acknowledgements}

PD acknowledges that Python and tikz code was written with the help of
Claude Opus 5. MG acknowledges fruitful discussions with Prof. A.
Baselga and funding under FCT (Portugal) Grant
FCT/Mobility/1419062306/2024-25.

\appendix

\section{Proofs}

If the vertex data assignment for a sampling graph $G_s$ is $D$, this
means that each vertex $v$ is assigned some data $D_v$ and
$|D_v|= \alpha_v$, the $\alpha$-diversity of the site $v$.

Recall the construction of $H(G_s)$. Let $d(v)$ denote the degree of a
vertex in $G_s$ and let $d'(v)$ denote its degree in $G_s'$. If
$d(v) \geq 2$ then $d'(v) = d(v)$ and if $d(v) = 1$ then $d'(v) =
2$. Define $c(v) = \binom{d'(v)}{2}$.

\subsection{Proof of Theorem~\ref{betass}}

The trace of $L_{H(G_s), H(D)}$ is the trace of $\delta^T\delta$,
where $\delta$ is the coboundary map. An edge $e$ in $H(G_s)$ with
label $D_e$ contributes $|D_e|$ rows to $\delta$ and each of these
rows contain exactly two non-zero entries: $+1$ and $-1$. Each row of
$\delta$ therefore contributes $1^2 + (-1)^2 = 2$ to the trace and
each edge contributes $2|D_e|$.

Each edge of $H(G_s)$ is generated by precisely one vertex of $G_s$
and each vertex $v$ generates a clique in $H(G_s)$ that contains
$c(v)$ edges. Thus, $v$ contributes $2c(v)\alpha_v$ to the trace of
$L_{H(G_s), H(D)}$. Summing over all vertices of $G_s$ gives
\[
E_1(H(G_s), H(D)) = 2\sum_{v \in V} c(v)\alpha_v.
\]
In the case of $D^{\max}$, $\alpha_v = \gamma$ for all $v$ and so
\[
E_1^{\max}(H(G_s), H(D)) = 2\gamma\sum_{v \in V} c(v).
\]
Hence,
\[
\beta_{ss}(G_s,D) = \frac{2\gamma\sum_{v \in V} c(v)}{2\sum_{v \in V} c(v)\alpha_v} = \frac{\gamma}{r_c}.
\]
 
\subsection{Proof of Theorem~\ref{paths_reg}}

An immediate corollary of Theorem~\ref{betass} is the fact that if
\(c(v)\) is constant for all vertices \(v\), then
\(\beta_{ss}(G_s,D) = \beta\) because $r_c$ reduces to the average
$\alpha$-diversity:
\[
r_c = \frac{\sum_{v \in V} c(v)\alpha_v}{\sum_{v \in V} c(v)} = \frac{\sum_{v \in V} \alpha_v}{\sum_{v \in V} 1} = \frac{1}{n}\sum_{v \in V} \alpha_v.
\]

In a path graph, $c(v) = 2$ for all vertices $v$ and in a $k$-regular
graph, $k \geq 2$, $c(v) = \binom{k}{2}$ for all $v$. Thus,
$\beta_{ss} = \beta$ in both cases. Note that a $1$-regular graph is a
path graph containing a single edge.

\bibliographystyle {plainnat}
\bibliography{Beta.bib}

\end{document}